\documentstyle[psfig]{article}
\begin{document}
{\baselineskip12pt
\vspace*{0.6cm}
\begin{center}
NUCLEAR HALO EFFECTS IN NEUTRON--CAPTURE REACTIONS OF ASTROPHYSICAL INTEREST
\end{center}
\vspace{1mm}
\begin{center}
H.~Oberhummer\footnote
{Tel.:~+43/1/58801/5574, E-mail:~ohu@ds1.kph.tuwien.ac.at, FAX:~+43/1/5864203},
H.~Herndl and R.~Hofinger\\
Institut f\"ur Kernphysik, TU Wien,\\
Wiedner Hauptstr.~8--10, A--1040 Vienna, Austria\\
\vspace{3mm}
Y.~Yamamoto\\
Department of Physics, Tokyo Metropolitan University,\\
1--1, Minami--osawa, Hachioji, Tokyo, 192--03, Japan
\end{center}}

{\bf Abstract:\/} The halo effect in the final states of astrophysically
relevant direct neutron--capture reactions by neutron--rich nuclei
is discussed. As an example, we calculate the cross sections
for $^{18}$O(n,$\gamma$)$^{19}$O at thermonuclear and thermal
energies.
\section{Introduction}
The neutron halo, one of the most prominent recent discoveries
in nuclear physics \cite{tan85,han87,tan91,zhu93,ris94,tan95},
is characteristic for neutron--rich nuclei. A halo effect
can also be observed in direct neutron--capture reactions leading
to final states with a loosely bound neutron~\cite{ots94}.
This halo effect can be important for astrophysically
relevant neutron--capture reactions by neutron--rich nuclei
in the $\alpha$-- and r--process occurring in supernovae
as well as in inhomogenous big--bang scenarios.
\section{Direct capture and nuclear halo effects}
In astrophysically relevant nuclear reactions two opposite
reaction mechanisms are of importance, compound--nucleus
formation and direct reactions. At the low reaction energies occurring in
primordial and
stellar nucleosynthesis the direct mechanism often cannot
be neglected and can even be dominant. The reason for
this behavior is that only a few levels exist for low
excitations of the compound nucleus. For instance, this is the
case for neutron capture by neutron--rich
nuclei. The importance of direct capture has
already been demonstrated for magic nuclei and
at the border of the region of stability (e.g., \cite{ohu96} for neutron capture 
by  neutron--rich nuclei).

The projectile--energy dependent factors in the Direct--Capture (DC) cross--section
$\sigma_{\rm DC}$ for an electric
dipole (E1) transition are given by~\cite{kim81,kim87,moh93,kra95}:
\begin{equation}
\sigma^{\rm E1}_{\rm DC} \propto  \frac{E_{\gamma}^3}{k}
\left| \int dr\: r^2 R_{\ell_{\rm f}}(r) \:
{\cal O}^{{\rm E1}}(r) \: \chi_{\ell_{\rm i}} (kr) \right|^2 \quad .
\end{equation}
In this expression the cross section is proportional to the square of
the radial overlap integral (direct--capture integral).
The photon energy is given by $E_{\gamma}$. 
The scattering wave
function in the entrance channel 
and the bound--state wave function in the exit channel
are given by $\chi_{\ell_{\rm i}}(kr)$ and $R_{\ell_{\rm f}}(r)$,
respectively. The kinetic energy $E$ in the entrance channel is related
to the wave number $k$ by $E = \hbar^2 k^2/(2 M)$, where $M$
is the reduced mass.
The radial part of the E1--transition operator is given in the
long wavelength approximation that is
appropriate for our low--energy 
(thermal and thermonuclear energies) case by  ${\cal{O}}^{{\rm E1}} \simeq r$.

For direct capture to weakly bound final states, the bound--state wave
function $R_{\ell j}(r)$ decreases only very slowly in the nuclear exterior,
so that the contributions come predominantly from far outside
the nuclear region, i.e., from the {\it nuclear halo\/}. For this asymptotic
region the
scattering and bound wave functions in Eq.~(1) can be approximated
by their asymptotic expressions neglecting the nuclear potential~\cite{ots94}
\begin{eqnarray}
\chi_{\ell_{\rm i}}(kr) & \propto & j_{\ell_{\rm i}}(kr)\\\nonumber
R_{\ell_{\rm f}}(r) & \propto & h_{\ell_{\rm f}}^{(+)}(i \mu r)  \quad ,
\end{eqnarray}
where $j_{\ell}$ and $h_{\ell}^{(+)}$ are the spherical
Bessel, and the Hankel function of the first kind, respectively.
The separation energy $S_{\rm n}$ in the exit channel
is related to the parameter $\mu$ by $S_{\rm n} = \hbar^2 \mu^2/(2 M)$.
\section{An example: $^{18}$O(n,$\gamma$)$^{19}$O}
\begin{figure}[htp]
\centerline{\psfig{file=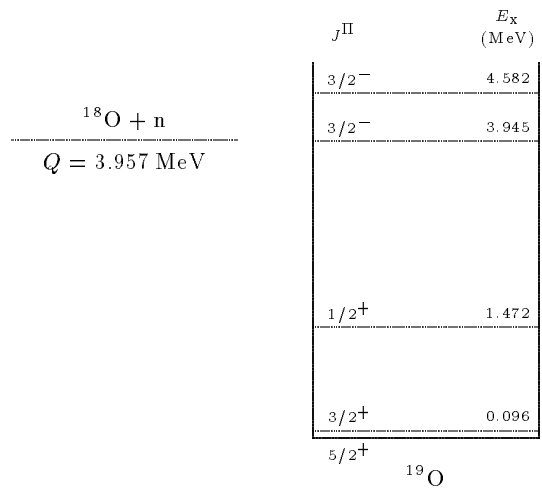,bbllx=193bp,bblly=422bp,bburx=420bp,bbury=661bp,clip=}}
\caption[]{\label{f1} Level scheme of $^{19}$O}
\end{figure}

\begin{figure}[htp]
\centerline{\psfig{file=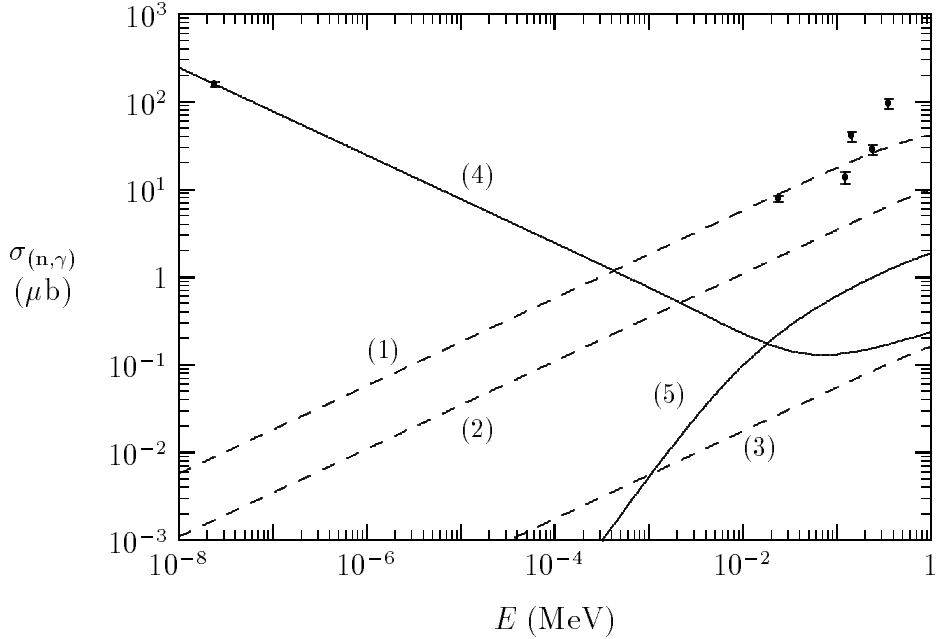,bbllx=122bp,bblly=306bp,bburx=505bp,bbury=590bp,clip=}}
\caption[]{\label{f2} Comparison
of direct--capture calculations with the
experimental data for the cross section of $^{18}$O(n,$\gamma$)$^{19}$O.
Explanation see text.}
\end{figure}

As an example we investigate the reaction $^{18}$O(n,$\gamma$)$^{19}$O.
The level
scheme of $^{19}$O is shown in Fig.~\ref{f1}.
The cross section for this reaction at thermonuclear energies
has been measured recently~\cite{mei96} and is also
known from experiment at thermal energies~\cite{sea92}. The
cross section has also been calculated using the
direct--capture model~\cite{mei96,gru95}. The
calculated direct--capture cross section is in excellent agreement
with the thermonuclear and thermal experimental cross section
(compare thick solid curve with the experimental data~\cite{mei96,sea92} in Fig.~\ref{f2}).
The calculations for the three main transitions to the positive--parity final states
in this work (the three broken curves labeled 1, 2 and 3 in Fig.~\ref{f2}) are the same as in
Ref.~\cite{gru95},
except that we used the spectroscopic factors extracted from
the $^{18}$O(d,p)$^{19}$O--reaction~\cite{sen74}. For the transitions
to the negative--parity $3/2^-$--state just below the
neutron threshold (the two solid
curves labeled 4 and 5 in Fig.~\ref{f2}), the spectroscopic
factor was adjusted to the experimental thermal cross
section~\cite{gru95}.

The halo effect shows up in the transition to the final $3/2^-$--state
in $^{19}$O, which is only bound by 12\,keV with respect to the
neutron threshold (Fig.~\ref{f1}). In this case
the main contributions come from a region of
about 75\,fm (s--wave) and 160\,fm (d--wave) at
thermal energies, and about 45\,fm (s--wave) and 85\,fm (d--wave) at thermonuclear
energies. The direct--capture
calculations result in two main transitions to this
state starting from an s-- and d--wave, respectively (the two solid
curves labeled 4 and 5 in Fig.~\ref{f2}). In this case the two
curves show a totally different behavior for below and above
a projectile energy that is equal to the neutron separation energy
$S_{\rm n} = 12$\,keV of the $3/2^-$--state.

This behavior can be readily described by the halo effect. For the
(s $\rightarrow$ p)--transition and (d $\rightarrow$ p)--transition,
we insert the specific expressions for the Bessel-- and Hankel functions
\begin{eqnarray}
j_{\ell_{\rm i}=0}(x) & = & \frac{\sin x}{x}\\\nonumber
j_{\ell_{\rm i}=2}(x) & = &
\left(\frac{3}{x^3} - \frac{1}{x}\right) \sin x - \frac{3}{x^2}{\cos x}\\\nonumber
h_{\ell_{\rm f}=1}^{(+)}(y) & = & \left(\frac{1}{y^2} - \frac{i}{y}\right) \exp (iy)
\quad ,
\end{eqnarray}
where $x=kr$ and $y=i\mu r$. The E1 direct--capture cross section is then given
by 
\begin{eqnarray}
\sigma^{\rm E1}_{\rm DC} ({\rm s} \rightarrow {\rm p}) & \propto & 
\frac{1}{\sqrt{E}} \frac{\left(E + 3S_{\rm n}\right)^2}{E + S_{\rm n}}\\\nonumber
\sigma^{\rm E1}_{\rm DC} ({\rm d} \rightarrow {\rm p}) & \propto &
\frac{E^{\frac{3}{2}}}{E+S_{\rm n}} \quad .
\end{eqnarray}

For $E \ll S_{\rm n}$ we recover the normal behavior at
low energies for an incoming s-- or d--wave:
\begin{eqnarray}
\sigma^{\rm E1}_{\rm DC} ({\rm s} \rightarrow {\rm p}) & \propto & 
\frac{1}{\sqrt{E}} \propto \frac{1}{v} \\\nonumber
\sigma^{\rm E1}_{\rm DC} ({\rm d} \rightarrow {\rm p}) & \propto &
E^{\frac{3}{2}} \propto v^3 \quad ,
\end{eqnarray}
where $v$ is the relative velocity in the entrance channel.

However, for $E \gg S_{\rm n}$ the energy behavior changes completely,
and we obtain:
\begin{eqnarray}
\sigma^{\rm E1}_{\rm DC} ({\rm s} \rightarrow {\rm p}) & \propto & 
\sqrt{E} \propto v \\\nonumber
\sigma^{\rm E1}_{\rm DC} ({\rm d} \rightarrow {\rm p}) & \propto & 
\sqrt{E} \propto v \quad .
\end{eqnarray}

Exactly this halo behavior is obtained for the transition to the
$^{19}$O(3.945\,MeV)--state, as can be seen easily from
the solid curves in Fig.~\ref{f2}. For low energies
the s--wave has an 1/v--behavior and the d--wave a v$^3$--behavior.
This is the normal energy dependence for low energies. However, at about 12\,keV this
behavior changes gradually in both cases to a v--behavior.
This effect has not been taken into account in previous publications~\cite{mei96,gru95}
of $^{18}$O(n,$\gamma$)$^{19}$O, where the s--wave was extrapolated
using an 1/v--behavior from the thermal cross section and the
d--wave has been neglected.

The contribution of the transition to the $^{19}$O(3.945\,MeV)--state
to the whole cross section of $^{18}$O(n,$\gamma$)$^{19}$O is about 5\,\%
in the astrophysically relevant range (10--250)\,keV (compare broken and solid
curves in Fig.~\ref{f2}). Therefore, the halo effect does not
play a large role in the reaction rate for $^{18}$O(n,$\gamma$)$^{19}$O
already given in Ref.~\cite{mei96}. However, from the astrophysical
point of view, when investigating neutron--capture reactions
further away from the region of stability this halo effect will be
of relevance in special cases, especially when approaching the drip lines.

{\it Acknowledgment:\/} This work was supported by the
Austrian Science Foundation (project S7307--AST) and by the \"Osterreichische
Nationalbank (project 5054).
\newpage
\end{document}